\documentclass[twocolumn,amsmath]{aastex701}

\shorttitle{Flux Variation of the Crab Pulsar}
\shortauthors{Augustine \& Chang}
\newcommand{\manuscriptfigure}[3]{%
  \IfFileExists{#1}{%
    \includegraphics[width=\textwidth,height=#2,keepaspectratio]{#1}%
  }{%
    \PackageWarning{CrabApJL}{Missing figure file #1; displaying a placeholder}%
    \fbox{\begin{minipage}[c][#3][c]{0.92\textwidth}
      \centering
      \textbf{Original figure file not supplied}\\[6pt]
      \texttt{\detokenize{#1}}\\[6pt]
      Upload this PDF to the project folder to display the figure.
    \end{minipage}}%
  }%
}

\begin{document}

\title{Flux Variation of the Crab Pulsar Observed Using Fermi LAT and NuSTAR}

\correspondingauthor{K.~A.~Augustine and H.-K.~Chang}

\author[0009-0008-7888-7584]{Koothodil Abhijith Augustine}
\affiliation{Institute of Astronomy, National Tsing Hua University, Hsinchu 300044, Taiwan}
\email[show]{augustinekoothodilabhijith@gmail.com}

\author[0000-0002-5617-3117]{Hsiang-Kuang Chang}
\affiliation{Institute of Astronomy, National Tsing Hua University, Hsinchu 300044, Taiwan}
\affiliation{Department of Physics, National Tsing Hua University, Hsinchu 300044, Taiwan}
\affiliation{Center for Theory-Computation-Data Science Research (CTCD), National Tsing Hua University, Hsinchu 300044, Taiwan}
\affiliation{Institute of Space Engineering, National Tsing Hua University, Hsinchu 300044, Taiwan}
\email[show]{hkchang@phys.nthu.edu.tw}

\makeatletter
\renewcommand{\@@email}{%
  augustinekoothodilabhijith@gmail.com\\
  hkchang@phys.nthu.edu.tw}
\makeatother

\begin{abstract}
The flux of pulsed emission from the Crab pulsar, just like most of other pulsars, is considered to be steady over time scales much larger than the pulsation period, except for a smooth, secular decrease related to the aging of pulsars. 
Such a steady assumption prevails in all theoretical investigations of pulsar emission models. 
However, here we report the discovery of stochastic variation in the flux and photon index of the Crab pulsar gamma-ray and hard X-ray emissions utilizing archival data of Fermi LAT (0.1 -- 100 GeV, spanning 17 years) and of NuSTAR (3.0 -- 79.0 keV, spanning 13 years) observations.
Furthermore, the flux variation is found to be anti-correlated with variation of the spectral photon index, that is, harder when brighter, in both gamma ray and hard X-ray bands.  
The origin of such variation is intriguing. 
Some possibilities are discussed.
\end{abstract}

\section{Introduction}\label{sec:introduction}
The Crab, including the nebula and the pulsar at the center,
was regarded as a steady emitter, a fundamental assumption that underpinned its use as a calibration source for X-ray and gamma-ray observatories. However, this paradigm was significantly challenged by the discovery of dramatic gamma-ray flares from the nebula by \textit{AGILE} and \textit{Fermi} LAT, revealing a far more dynamic and turbulent environment than previously thought \citep{2011Sci...331..739A, 2011Sci...331..736T, 2014RPPh...77f6901B}. Less dramatic, long-term variations in hard and soft X-ray fluxes were also reported recently \citep{2025ApJ...981..173A, 2025ApJ...988..105C}. These flares and fluctuations are attributed to the nebula, and the pulsar emission remains to be considered steady. In this work, we present spectral analysis results of the pulsed emission from the Crab pulsar using long-term archival data of Fermi LAT and NuSTAR observations to examine its flux stability. 

\section{Observations and Analysis Overview}\label{sec:overview}
We analyze monthly Fermi LAT data in the 0.1--100~GeV band and NuSTAR data in the 3--79~keV band. In both bands, we isolate pulsed emission using on-pulse intervals $\phi \in [0.00,0.52]\cup[0.87,1.00]$ and the off-pulse interval $\phi \in [0.52,0.87]$. For Fermi LAT, each month's off-pulse fit constrains the nebular and background components in the corresponding on-pulse analysis. For NuSTAR, on-pulse spectra are fitted with off-pulse spectra as background, and sequential observations are combined into 43 measurement epochs with at least 5~ks of exposure. Statistical and systematic uncertainties are included as described in Appendix~\ref{methods}, which provides the full data selection, timing, and spectral-analysis procedures.

\section{Results}\label{sec:results}
\begin{figure*}[tp]
    \centering
    \manuscriptfigure{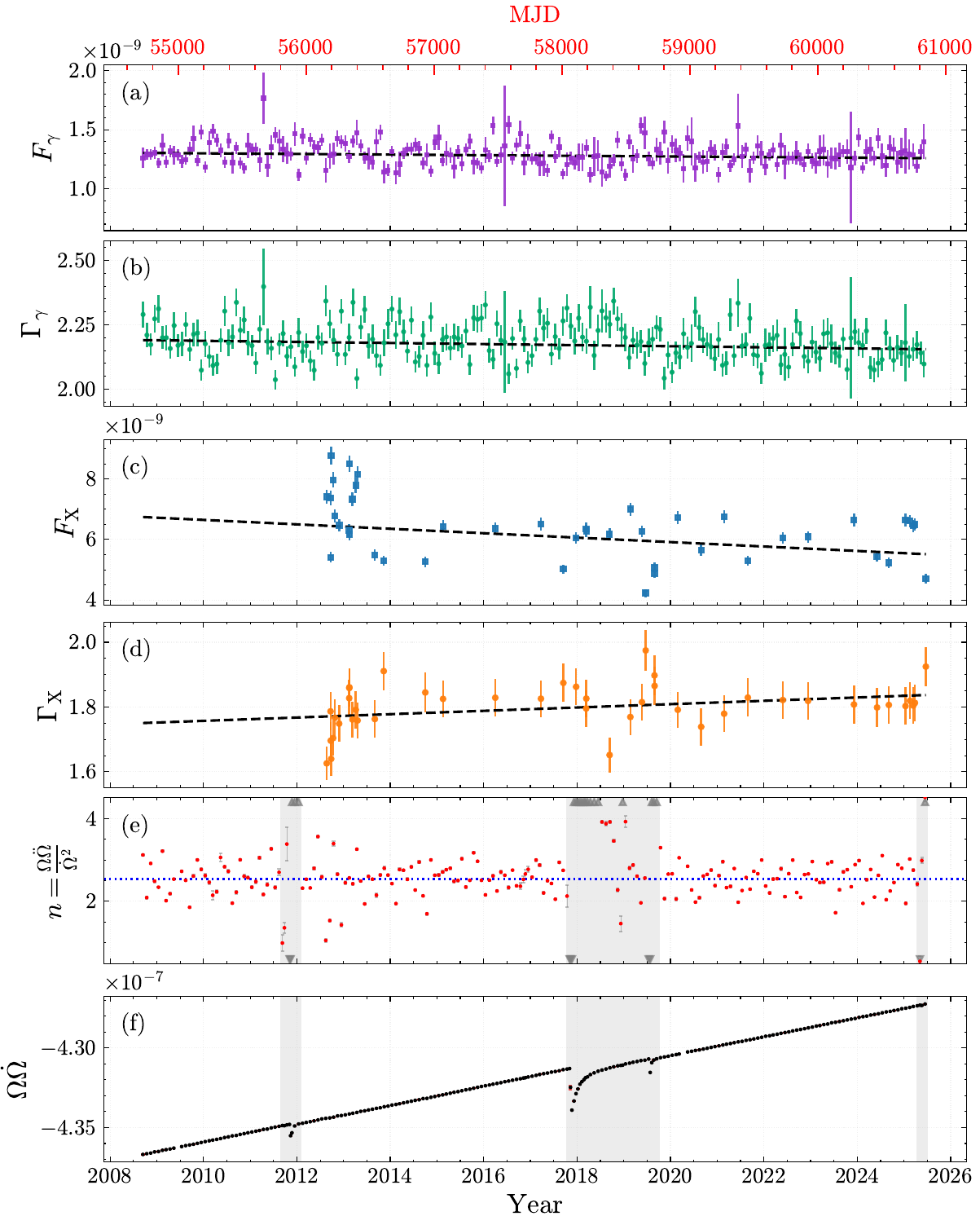}{0.70\textheight}{0.38\textheight}
    \caption
        {
    \textbf{Spectral- and timing-property evolution of the Crab pulsar.}   
    \textbf{(a)} Monthly Fermi LAT gamma-ray energy flux $F_{\gamma}$ (0.1--100 GeV) of the pulsed emission. 
    \textbf{(b)} Corresponding evolution of the Fermi LAT photon index $\Gamma_{\gamma}$.
    \textbf{(c)} NuSTAR X-ray energy flux $F_{\rm X}$ (3.0--79.0 keV) of the pulsed emission. 
    \textbf{(d)} Corresponding evolution of the NuSTAR X-ray photon index $\Gamma_{\rm X}$. 
    \textbf{(e)} Braking index, $n = \Omega \ddot{\Omega} / \dot{\Omega}^2$.
    Gray triangles at the top and the bottom in the shaded segments indicate values outside the display range of this figure panel.
    \textbf{(f)} Evolution of the product $\Omega\dot{\Omega}$, which is the rotational energy loss rate divided by moment of inertia. 
    In all panels, the bottom axis denotes the calendar year and the top axis of panel {\bf (a)} provides the Modified Julian Date (MJD). Dashed black lines in panels \textbf{(a)}--\textbf{(d)} represent ordinary least squares linear best fits to the data points. Vertical gray shaded segments in panels \textbf{(e)} and \textbf{(f)} mark the epochs and recovery windows of major glitches. All error bars denote $1\sigma$ uncertainties with both statistical and systematic uncertainties added in quadrature for Fermi LAT and NuSTAR observations.}
    \label{fig:lightcurve}
\end{figure*}

The long-term evolution of the Crab pulsar's pulsed emission is shown in Figure~\ref{fig:lightcurve}, 
where significant, stochastic variability can be seen in both the gamma-ray (Fermi LAT) and X-ray (NuSTAR) bands. 
All errors are $1 \sigma$ uncertainty and the statistical and systematic errors are both included. In the gamma-ray regime (MJD 54710 to 60842), the energy flux ($F_\gamma$) derived from monthly integrated data fluctuates around a mean of about  $(1.283 \pm 0.004) \times 10^{-9} $~erg~cm$^{-2}$~s$^{-1}$, with individual monthly excursions varying by a few percents up to about 20\%. 
The corresponding photon index ($\Gamma_\gamma$) behaves similarly with a mean of about $2.17 \pm 0.02$ and fluctuations up to about 10\%.

A linear fit over time for the Fermi LAT flux and photon index (shown as dashed black lines in Figure~\ref{fig:lightcurve}) yields best-fit slopes of $(-2.65 \pm 0.94) \times 10^{-12}$~erg~cm$^{-2}$~s$^{-1}$~yr$^{-1}$ and $(-1.92 \pm 0.59) \times 10^{-3}$~yr$^{-1}$, respectively. 
The best-fit statistics using reduced chi-square ($\chi^{2}_{\nu} = 1.9$, $p = 2.6 \times 10^{-14}$ for $F_\gamma$; $\chi^{2}_{\nu} = 2.7$, $p = 6.8 \times 10^{-32}$ for $\Gamma_\gamma$) indicate massive deviations from a simple linear evolution.

In the X-ray band, with very small error bars derived from the NuSTAR data, the variability is much more clearer than that in the gamma-ray band. The mean of the NuSTAR flux is $(6.086 \pm 0.002) \times 10^{-9} $~erg~cm$^{-2}$~s$^{-1}$ and the photon index is $1.80 \pm 0.05$.
Linear fits to the X-ray flux and photon index yield slopes of $(-7.95 \pm 0.69) \times 10^{-11}$~erg~cm$^{-2}$~s$^{-1}$~yr$^{-1}$ for $F_{\rm X}$ and $(+5.37 \pm 1.95) \times 10^{-3}$~yr$^{-1}$ for $\Gamma_{\rm X}$, respectively, and is shown with dashed black lines in Figure~\ref{fig:lightcurve}. 
The exceptionally high reduced chi-square value ($\chi^{2}_{\nu} = 22.5$, $p \sim 0$ for $F_{\rm X}$ and $\chi^{2}_{\nu} = 1.4$ for $\Gamma_{\rm X}$, with $p = 4.7 \times 10^{-2}$) formally reject the null hypothesis of steady emission.

The X-ray variation is particularly significant in the pre-2014 NuSTAR observations. To our knowledge, there is no study reporting any anomalous issues about NuSTAR data before 2014.
Though, we also examined the results excluding the pre-2014 data. In this case, the mean flux of the NuSTAR data is $(5.913 \pm 0.002) \times 10^{-9} $~erg~cm$^{-2}$~s$^{-1}$ and the photon index is $1.82 \pm 0.06$. Linear fits to the X-ray flux and photon index yield slopes of $(-4.85 \pm 0.78) \times 10^{-11}$~erg~cm$^{-2}$~s$^{-1}$~yr$^{-1}$ for $F_{\rm X}$ and $(+1.30 \pm 2.3) \times 10^{-3}$~yr$^{-1}$ for $\Gamma_{\rm X}$, respectively, with the reduced chi-square value $\chi^{2}_{\nu} = 18.83$, ($p \sim 0$) for $F_{\rm X}$ and $\chi^{2}_{\nu} = 1.01$ ($p = 0.45$) for $\Gamma_{\rm X}$. Even if the pre-2014 variability is excluded, the NuSTAR flux is still highly variable.

\begin{figure*}[tp]
    \centering
    \manuscriptfigure{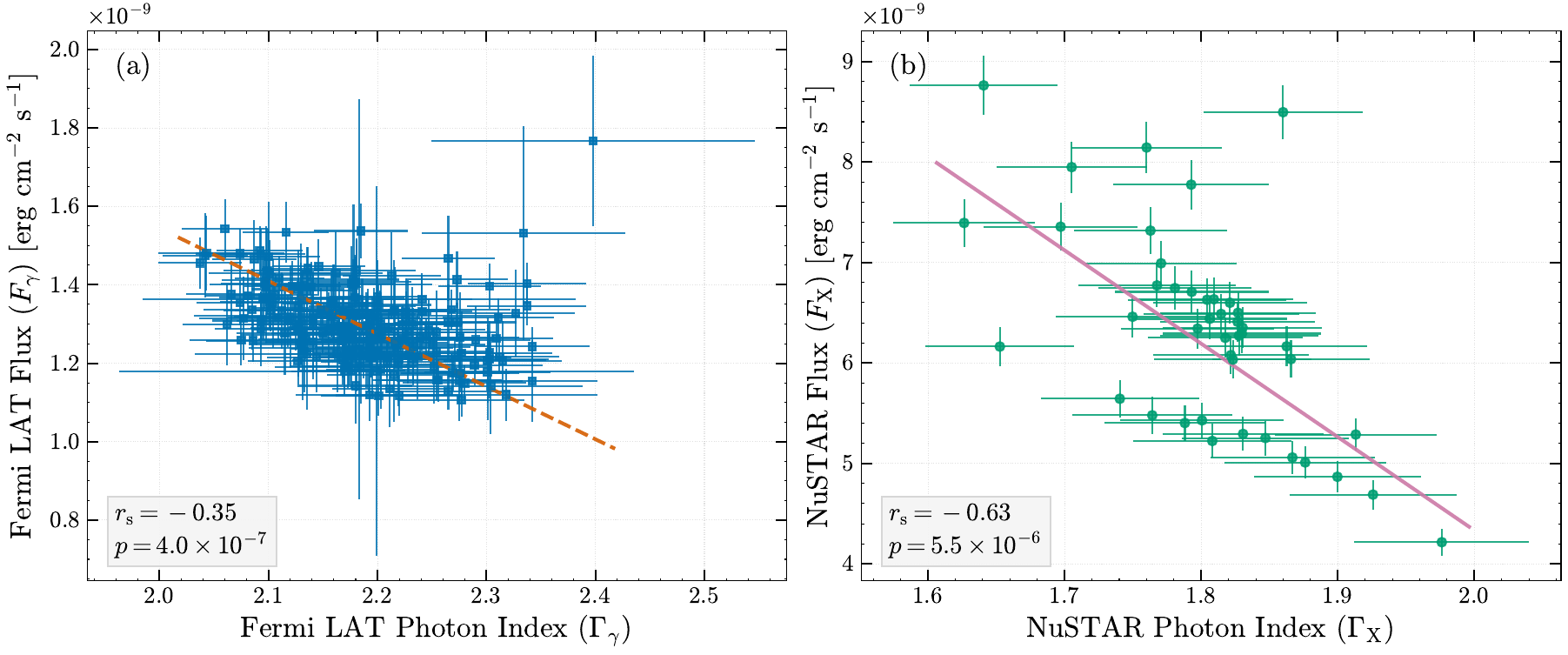}{0.48\textheight}{0.18\textheight} 
    \caption{
    \textbf{Spectral correlations of the Crab pulsar in gamma-ray and X-ray bands.}
    \textbf{(a)} Correlation between the Crab pulsar's Fermi LAT flux (0.1--100 GeV) and the gamma-ray photon index. The dashed orange line indicates the best-fit linear model obtained via Orthogonal Distance Regression (ODR), which accounts for uncertainties in both variables. A highly significant negative correlation is present ($r_{\rm s} = -0.35$, $p = 4.0 \times 10^{-7}$). 
    \textbf{(b)} Correlation between the NuSTAR flux (3.0--79.0 keV) and the X-ray photon index. The solid red line indicates the best-fit linear model obtained via ODR regression. A strong negative correlation is observed ($r_{\rm s} = -0.63$, $p = 5.5 \times 10^{-6}$). 
    Data points are shown with $1\sigma$ uncertainties. Both distinct energy regimes demonstrate a consistent ``harder-when-brighter'' spectral evolution.
    }
    \label{fig:anticorrelation}
\end{figure*}

The aforementioned statistics confirm the presence of highly significant, stochastic variability in the Crab pulsar's emission on top of underlying secular trends.
Another critical finding of this study is anti-correlations between the energy flux and the photon index, in both gamma-ray and X-ray bands, which are shown in Figure~\ref{fig:anticorrelation}. 
We find strong negative correlations in both energy bands with the Spearman rank-order correlation coefficient $r_{\rm s} = -0.35$ ($p = 4.0 \times 10^{-7}$) and $r_{\rm s} = -0.63$ ($p = 5.5 \times 10^{-6}$), respectively, confirming a broadband ``harder-when-brighter'' behavior. 
On the other hand, no clear correlation is found between the gamma-ray and X-ray fluxes, by comparing each NuSTAR flux and the Fermi LAT flux of the corresponding month. 
NuSTAR observations are pointed ones with  certain exposure times at different epochs,
while that of Fermi LAT are monthly integrated data. It indicates that the variability also exists at a time scale shorter than a month.
Unfortunately Fermi LAT data does not provide
measurements precise enough (that is, with small enough error bars) on a daily basis. 

\begin{figure*}[tp]
    \centering
    \manuscriptfigure{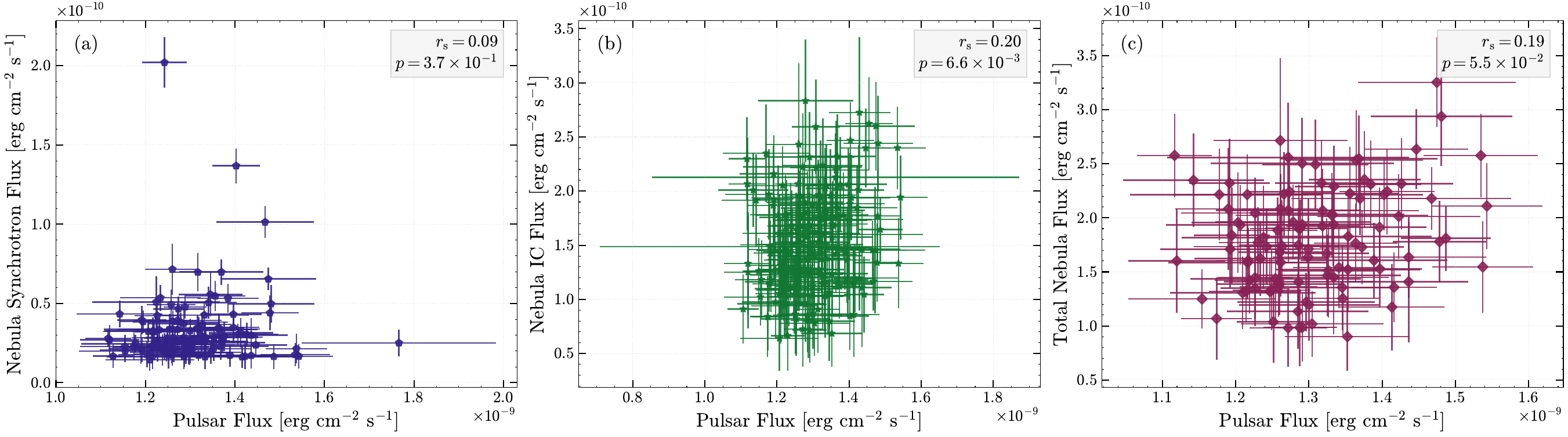}{0.48\textheight}{0.18\textheight} 
    \caption{
    \textbf{Comparison of nebular flux with pulsar flux for Fermi LAT Crab data.}
    \textbf{(a)} Correlation between nebula synchrotron flux and pulsar flux. No significant correlation is obtained between them with $r_{\rm s} = +0.09$, $p = 3.70 \times 10^{-1}$.
    \textbf{(b)} Correlation between nebula inverse Compton flux and pulsar flux. No significant correlation is obtained between them with $r_{\rm s} = +0.18$, $p = 6.57 \times 10^{-3}$.
    \textbf{(c)} Correlation between total nebular flux and pulsar flux. No significant correlation is obtained between them with $r_{\rm s} = +0.19$, $p = 5.51 \times 10^{-2}$.
    }
    \label{fig:correlation}
\end{figure*}

To examine the possibility that the pulsed flux variation might be due to possibly improper separation of the fluxes between that from the nebula and the pulsed component, the correlation between pulsar flux and nebula flux in the Fermi LAT data is studied and is shown in Figure~\ref{fig:correlation}. From a total of 202 months, the number of valid detections for the synchrotron part of the nebula spectrum is 111, while that for the inverse Compton component is 189. The detection validity is examined by checking if the detection significance is above 2 sigma. Neither the synchrotron flux nor the inverse Compton flux show a strong correlation with pulsed flux with Spearman rank-order correlation coefficient $r_{\rm s} = +0.09$ ($p = 0.37$) and $r_{\rm s} = +0.18$ ($p = 6.57 \times 10^{-3}$), respectively. That for the total nebular flux has a Spearman rank-order correlation coefficient of $r_{\rm s} = +0.19$ ($p = 5.51 \times 10^{-2}$) against the pulsar flux showing that there is no statistically significant correlation between them. This assures us that the nebular component has been properly separated from the pulsed component.

\section{Discussion}\label{sec:discussion}
 
The variation of pulsed emission from pulsars, as reported in this paper for the Crab pulsar, is not expected by any theoretical models in the literature. In this paper we present the result of our analysis, taking into account both statistical and systematic errors.
The variation is significant at a monthly time scale or shorter. 
The anti-correlation between fluxes and photon indices also provides another supporting evidence for the variation.
Random fluctuations will not result in such correlations.

To understand the origin of such variation relies on our understanding of pulsars' high-energy emission mechanism.
In addition to enhancing the truthfulness of the variation reported in this paper, the anti-correlation between fluxes and photon indices clearly provides some constraints on models for pulsars' particle acceleration and radiation mechanisms. 
It is generally proposed that X-ray emission from the Crab pulsar is due to synchrotron radiation and the gamma-ray one is due to synchrotron-self-Compton scattering \citep{2007ApJ...670..677T, 2017ApJ...834....4T, 2021ApJ...923..194H}. 
The emission site may be in outer regions in the magnetosphere or in the wind zone \citep{2022ARA&A..60..495P}.
There are two ways to increase the radiation flux. One is to increase the number of radiating particles (in the corresponding, relevant energy range) and the other is to increase the magnetic field strength or particles' pitch angles.
These two possibilities, however, do not come automatically with a harder spectrum.
Stronger magnetic fields may result in more energetic particles via electric acceleration in the magnetosphere (outer-gap and slot gap models) and associated pair-production cascade.
However, the effect is only to increase the number of particles in the corresponding energy range but not to make the energy distribution of the radiating particles harder, which is needed to have a harder radiation spectrum. 

It is worth investigating whether there is a certain mechanism that links the energy distribution of radiating electron-positron pairs with their pitch angles in such a way that a harder distribution tends to come with larger pitch angles. This is apparently related to how these pairs are created.
Another possibility is that, in the current sheet acceleration scenario in the wind zone, the energy distribution of the particles may be softer when the pair-multiplicity
is higher for a given magnetic field strength.
This softer distribution will result in a softer synchrotron emission and, because the emission power depends on the number of particles linearly but on the particle energy quadratically, a higher pair-multiplicity leads to a lower emission power. The anti-correlation reported in this paper, that is, harder when brighter, could thus be explained.  
However, even when the anti-correlation can be explained, the origin of the variation remains mysterious.

One possible cause for the flux variation (and maybe also the aforementioned anti-correlation) might be rooted in the origin of the primary electron-positron pairs that are assumed to be created and accelerated in the magnetic polar gap region. These pairs further make secondary pairs above the polar cap or in the outer magnetosphere (outer gaps or slot gaps) via pair-production cascade to produce an ample amount of pairs to contribute to observed radiations. The energy distribution and the total amount of secondary pairs depend on those of primary pairs, which are mainly related to the so-called Goldreich-Julian charge density and the available electric voltage in the polar gap region, which in turn depend on pulsar spin angular velocity, magnetic field strength near the magnetic pole and the magnetic inclination angle \citep{1975ApJ...196...51R}.

Therefore, one not impossible conjecture is that the observed flux and photon index fluctuations might be due to possible stochastic variations in neutron star's magnetic field strength at the pole and magnetic inclination angle.  To examine this possibility further, we also track the evolution of the braking index of the Crab pulsar, as shown in panel (e) in Figure~\ref{fig:lightcurve}. 
Assuming magnetic dipole braking and a spin-down law in the form of $\dot{\Omega}=-K\Omega^3$, where $K$, which can be time-dependent, is a function of magnetic dipole moment $\mu$, magnetic inclination angle $\alpha$ and neutron star's moment of inertia $I$, the pulsar braking index $n$ is defined as \citep{2015MNRAS.446..857L}:
\begin{equation}
n\equiv\frac{\Omega\ddot{\Omega}}{\dot{\Omega}^2} = 3 + \frac{\Omega}{\dot{\Omega}}\frac{\dot{K}}{K}\,\,\, .
\end{equation}
As shown in Figure~\ref{fig:lightcurve},
the Crab pulsar's braking index deviates from 3 significantly. It indicates that $\dot{K}$ is non-zero. Excluding glitch epochs (the shaded segments in panels (e) and (f) in Figure~\ref{fig:lightcurve}), this braking index has an average value of 2.54 and also fluctuates  between about 1.5 and 3.5.

From the average value of the braking index, one can infer that $\dot{K}/K$ has an average value of about $5.8\times 10^{-12}$ s$^{-1}$ (adopting $\nu=\frac{\Omega}{2\pi}=29.947$ Hz and $\dot\nu=-3.7754\times 10^{-10}$ Hz s$^{-1}$).
This has been explained with the secular increase in $\alpha$ inferred from a long-term observation of the Crab's radio pulse profile \citep{2013Sci...342..598L, 2015MNRAS.446..857L} by noting that, under the magnetic dipole braking assumption,
\begin{equation}
    \frac{\dot{K}}{K}=-\frac{\dot{I}}{I}+2\frac{\dot{\mu}}{\mu}+2\frac{\dot{\alpha}}{\tan\alpha}
\end{equation}
and assuming no contribution from the $\dot{I}$ and $\dot{\mu}$ terms.
This explanation accounts for the braking index's deviation from 3, but not for its fluctuations, which requires the
$\dot{K}/K$ term to vary roughly from $-6.3\times 10^{-12}$ s$^{-1}$ (for $n=3.5$) to $+1.9\times 10^{-11}$ s$^{-1}$ (for $n=1.5$).

It is not clear how large a $\dot\mu$ and a $\dot\alpha$ are needed to result in changes of the energy distribution and amount of primary electron-positron pairs that could explain the observed variation in gamma-ray and X-ray fluxes and photon indices.  
The fractional changing rate of the gamma-ray flux is about a few percents per month, which is roughly $10^{-8}$ s$^{-1}$ (assuming 3\% per month). 
The fractional changing rate in the pulsar spin-down energy-loss rate due to magnetic dipole braking, 
which is 
%$\frac{\ddot{E}}{\dot{E}}=2\frac{\dot{\mu}}{\mu}+2\frac{\dot{\alpha}}{\tan\alpha}+4\frac{\dot{\Omega}}{\Omega}$,
$\ddot{E}/\dot{E}=2\dot{\mu}/\mu+2\dot{\alpha}/\tan\alpha+4\dot{\Omega}/\Omega$,
could be of the same order, 
if a constant efficiency of turning the spin-down power into gamma- and X-ray luminosity is assumed.
In such a case, because $\dot{\Omega}/\Omega$ is of order $10^{-11}$ s$^{-1}$,
the $2\dot{\mu}/\mu+2\dot{\alpha}/\tan\alpha$ 
term should be of order $10^{-8}$ s$^{-1}$, 
which is about thousand times larger than what is needed to account for the braking index deviation (from 3) or fluctuations.

On the other hand, changes in the magnetic dipole moment or the inclination angle, which may be due to interactions among superfluid vortices and superconducting magnetic flux tubes, may also change the moment of inertia accordingly, because of accompanying plasma motions and changes in the mass distribution in neutron stars \citep{Lander:2024kye, 2026EPJWC.35602015G, 2026ApJ...998..319B}. The rotation energy changing rate, $\dot{E}=I\Omega\dot\Omega$, as shown in panel (f) in Figure~\ref{fig:lightcurve}, changes smoothly except for glitch epochs if the moment of inertia $I$ is dropped out. It suggests that, under the magnetic dipole braking assumption, the moment of inertia $I$ may adjust itself in accordance with the change in $\mu$ and $\alpha$. 
Therefore, a tempting but tricky scenario is that the $\dot I/I$ term and at least one of the $\dot\mu/\mu$ and $\dot\alpha/\tan\alpha$ terms vary at the level of order $10^{-8}$ s$^{-1}$ synchronously but with a tiny, stochastically varying difference thousands times smaller.
The requirement of this fine tuning, however, makes this scenario less favorable.

The variation in the efficiency of pulsar high-energy emissions might also be a possible cause of the flux variation. 
However, the efficiency, which is the ratio of luminosity to spin-down power, usually depends on factors like $\Omega$, $\dot\Omega$, $\mu$, $\alpha$, and $I$ only. 
To have a fractional changing rate of $10^{-8}$ s$^{-1}$ in the efficiency requires a similar changing rate in at least one of those factors.
It in turn suggests such a rate for $\mu$, $\alpha$, and $I$.
This will again lead to the tricky scenario discussed above because of the constraint imposed by the observed braking index. 
On the other hand, if the efficiency depends on the surface temperature in some way, it may provide a solution. A few percents change in the surface temperature at a monthly time scale or shorter seems possible if electric currents returning from the magnetosphere to the surface are also stochastic in nature.
The temperature dependence of pulsars' non-thermal X-ray and gamma-ray emissions was reported in \citet{2023MNRAS.520.4068C} and \citet{2026MNRAS.546ag142A}.

Observed radio fluxes from pulsars may vary due to interstellar scintillation, but their radio luminosities are found to be steady \citep{2000ApJ...539..300S} within the uncertainty so far.
The gamma-ray flux from the Vela pulsar has been reported to be variable based on COS-B observations \citep{1988A&A...204..117G}, but the variability was only marginally found in later observations with EGRET \citep{1995ApJ...450..791R, 2003ApJ...597..615N} and totally undetected in Fermi LAT observations \citep{2010ApJ...713..154A}. 
A more detailed study with new and more data on pulsar flux variability is desired.

\section{Conclusions}\label{sec:conclusions}
In all the earlier studies, the Crab pulsar was always reported to be a steady emitter. 
In this paper, with long-term Fermi LAT and NuSTAR data, we report the discovery of variation in the flux and photon index of pulsed gamma-ray and X-ray emissions from the Crab pulsar and the anti-correlations between them.
This discovery provides an opportunity for us to understand neutron star physics better. 
The actual mechanism
responsible for these variations and correlations  requires more detailed, more quantitative modeling. 
It is also tempting to investigate whether these variations are also present in radio luminosities or in other pulsars, if required monitoring observations are feasible.

\begin{acknowledgments}
K.~A.~A. and H.-K.~C. are both funded by the National Science and Technology Council of Taiwan under grants NSTC 114-2112-M-007-042 and NSTC 115-2112-M-007-029-MY3.
\end{acknowledgments}

\begin{contribution}
Koothodil Abhijith Augustine and Hsiang-Kuang Chang contributed equally to this work.
\end{contribution}

\section*{Data and Code Availability}
The data underlying this article are publicly available in the following repositories:
\begin{itemize}
    \item \textbf{NuSTAR Data:} Available through the NASA HEASARC archive at \url{https://heasarc.gsfc.nasa.gov/cgi-bin/W3Browse/w3browse.pl}.
    \item \textbf{Fermi LAT Data:} Available through the Fermi Science Support Center (FSSC) at \url{https://fermi.gsfc.nasa.gov/cgi-bin/ssc/LAT/LATDataQuery.cgi}.
    \item \textbf{Crab Pulsar Ephemeris:} The Jodrell Bank Crab Pulsar Monthly Ephemeris used for phase assignment is publicly available at \url{https://www.jb.man.ac.uk/pulsar/crab.html}.
\end{itemize}
Software packages utilized for data reduction, including \textsc{heasoft} and \textsc{Fermitools}, are openly distributed by NASA's HEASARC and FSSC, respectively.

\facilities{Fermi (LAT), NuSTAR}

\bibliography{sample701}
\bibliographystyle{aasjournalv7}
    
\clearpage
\appendix
\section{Detailed Data Analysis}\label{methods}

\subsection{Fermi LAT}

Data collected by the Large Area Telescope (LAT) aboard the Fermi Gamma-Ray Space Telescope mission were utilized to analyze the pulsed emission of the Crab pulsar across the $100$~MeV to $100$~GeV energy regime. The LAT is a pair-conversion telescope with a 2.4-steradian field of view that surveys the entire sky every 3.2~hours \citep{2009ApJ...697.1071A}. Our data set covers a temporal baseline of approximately 17 years, spanning the period from September 2, 2008, to June 16, 2025 (MJD $54710$--$60842$). We defined a $15^{\circ} \times 15^{\circ}$ Region of Interest (ROI) centered on the J2000 coordinates of the Crab pulsar ($\mathrm{RA} = 83.63^{\circ}$, $\mathrm{Dec} = 22.01^{\circ}$).

The analysis was performed using \texttt{Fermitools} (v2.2.0) wrapped by the \texttt{Fermipy} (v1.4.0) Python package, utilizing the \texttt{P8R3\_SOURCE\_V3} instrument response functions (IRFs). We applied the \texttt{evclass=128} selection to isolate the ``Source'' event class, complemented by an \texttt{evtype=3} filter to include both front and back conversion events. To mitigate background contamination originating from the atmospheric limb of the Earth, a strict cut in the zenith angle of $105^{\circ}$ was enforced. The filter expression used for data quality was \texttt{(DATA\_QUAL>0) \&\& (LAT\_CONFIG==1) \&\& ABS(ROCK\_ANGLE)<52}. 
Barycentric time correction was conducted with {\tt gtbary}.
A binned maximum likelihood analysis was used with a fine spatial binning of $0.01^\circ$ per pixel and 10 bins per energy decade. To maintain spectral precision at high energies, we enabled energy dispersion (\texttt{edisp}) corrections for all components in the source model, excluding isotropic diffuse emission.

\subsubsection{Phase-Resolved Analysis}

We utilized a phase-resolved gating technique to isolate the pulsed signal from the emission of the surrounding Crab Nebula. Rotation phases were assigned to each photon using \texttt{PINT} \citep{2021ApJ...911...45L} with the monthly ephemeris timing solutions provided by the Jodrell Bank Observatory \citep{2015MNRAS.446..857L}. 

We first performed a maximum likelihood analysis on the ``off-pulse'' interval ($\phi \in [0.52, 0.87]$), where the pulsed emission is negligible. The Galactic diffuse emission was modeled using the \texttt{gll\_iem\_v07.fits} template, and the isotropic component was accounted for using the \texttt{iso\_P8R3\_SOURCE\_V3\_v1.txt} template. To accurately capture the broadband steady emission, we modeled the Crab Nebula utilizing two independent Log-Parabola spectra to represent the synchrotron and inverse Compton components, respectively. All 4FGL-DR4 catalog sources within the ROI were also included in the model.

When analyzing the ``on-pulse'' intervals ($\phi \in [0.0, 0.52] \cup [0.87, 1.0]$), the best-fit parameters for the background sources and the nebula were frozen at their off-pulse values for each individual month. By utilizing the steady emission as a fixed baseline, we robustly extracted the flux and photon index of the pulsed component. 
To quantify systematic uncertainties, we performed tests on the optimal likelihood fit to account for the limits of instrument calibration and background modeling. We evaluated the uncertainty of the interstellar emission model by artificially shifting the normalization of the Galactic diffuse template by $\pm 6\%$ \citep{2023ApJ...958..191S}. Separately, we accounted for the energy-dependent systematic uncertainty of the effective area ( $5\%$ at $100$~MeV, $3\%$ at $1$~GeV, and $10\%$ above $10$~GeV) by re-evaluating the fit using the dedicated bracketed IRFs \footnote{\url{https://fermi.gsfc.nasa.gov/ssc/data/analysis/scitools/Aeff_Systematics.html}}. In the analysis, we set the \texttt{edisp=True} to account for the energy dispersion. The maximum absolute flux shifts resulting from the diffuse and IRF perturbations were combined in quadrature to establish the final systematic uncertainty envelope.

\subsubsection{Pulsed Spectra}

The pulsar spectrum was modeled using the \texttt{PLSuperExpCutoff4} (PLEC4) functional form, which was uniformly employed for high-energy pulsars in the Fermi LAT Third Pulsar Catalog (3PC) \citep{2023ApJ...958..191S}. The model is defined as:
\begin{equation}
    \frac{dN}{dE} = N_0 \left( \frac{E}{E_0} \right)^{-\gamma_0 + \frac{d}{b}} \exp \left[ \frac{d}{b^2} \left( 1 - \left( \frac{E}{E_0} \right)^b \right) \right] \, ,
    \label{eq:pulsed_spectrum}
\end{equation}
where $N_0$ is the prefactor, $E_0$ is the scale energy, $d$ (\texttt{ExpfactorS}) represents the curvature strength, and $b$ (\texttt{Index2}) is the exponential index. For the Crab pulsar, we fixed $b$ at $0.52$ to reflect the stable sub-exponential cutoff shape and $E_0$ at $1304$~MeV to stabilize the spectra observed in long-term averages. These values are obtained from the DR4 of the 4FGL catalog \citep{2023arXiv230712546B}. The photon index in this parameterization, which we denote as $\Gamma_{\gamma}$ throughout our analysis, corresponds to the term $\Gamma_\gamma = \gamma_0 -\frac{d}{b}$.

\subsubsection{Nebula Spectra}

To capture the broadband spectral shape of the Crab Nebula, the emission was modeled using two Log-Parabola functions corresponding to the synchrotron and inverse Compton emissions. The Log-Parabola model is defined as:
\begin{equation}
    \frac{dN}{dE} = N_0 \left( \frac{E}{E_0} \right)^{-\alpha - \beta \ln\left(\frac{E}{E_0}\right)} \, ,
    \label{eq:nebula_spectrum}
\end{equation}
where $N_0$ is the normalization factor, $E_0$ is the fixed reference energy, $\alpha$ is the photon index at $E_0$, and $\beta$ is the curvature parameter signifying the degree of spectral bending. Out of 202 total observations (1 for each month), we successfully detected 111 for synchrotron and 189 for inverse Compton observations.

\subsection{NuSTAR}

\subsubsection{Data Analysis}

We analyzed X-ray ($3$--$79$~keV) observations of the Crab pulsar utilizing the Nuclear Spectroscopic Telescope Array (NuSTAR) data \citep{2013ApJ...770..103H}. The data were processed using the NuSTAR Data Analysis Software (NuSTARDAS), distributed as part of the HEASoft package. Initial calibration and screening of the raw Level 1 telemetry were performed using the \texttt{nupipeline} tool. We applied standard depth corrections, excluded periods of high instrumental background associated with the South Atlantic Anomaly (SAA), and applied the latest instrumental calibration database (CALDB) to produce cleaned Level 2 event files for both focal plane modules (FPMA and FPMB).

To perform precise phase-resolved spectroscopy and isolate the pulsed emission, it is necessary to correct the photon arrival times for the motion of the spacecraft and the Earth. The arrival time of each photon was translated to the Solar System Barycenter (SSB) using the \texttt{barycorr} tool. We adopted the Jet Propulsion Laboratory (JPL) DE430 solar system ephemeris and the precise J2000 equatorial coordinates of the Crab pulsar (R.A. = $83.63322^\circ$, Decl. = $22.01446^\circ$, ICRS reference frame).

Following the barycentric correction, an absolute rotational phase ($\phi \in [0.0, 1.0]$) was assigned to each X-ray event using the \texttt{photonphase} utility from PINT. The events were folded utilizing a contemporary timing ephemeris taken from Jodrell Bank Observatory \citep{1993MNRAS.265.1003L}, accounting for the pulsar's rotational frequency ($\Omega$) and its first derivative ($\dot{\Omega}$), interpolated to the exact epoch of each NuSTAR observation.

Because our analysis focuses specifically on the pulsed emission, we defined strict temporal phase cuts to the X-ray light curve of the Crab pulsar \citep{2015ApJS..220....8M}. Photons arriving in the ``on-pulse'' phase intervals $\phi \in [0.00, 0.52]$ and $\phi \in [0.87, 1.00]$ were combined.
The off-pulse phase is defined to be $\phi \in [0.52, 0.87]$. These temporal phase cuts were applied to Level-2 event lists utilizing the \texttt{fselect} task.

\subsubsection{Spectral Analysis}

Standard PHA files for the on-pulse and off-pulse phase intervals were extracted using the \texttt{nuproducts} tool, simultaneously generating the corresponding Redistribution Matrix Files (RMFs) and Ancillary Response Files (ARFs). We selected a circular region of $60 ''$ centered around the equatorial coordinates of the Crab (RA = $83.63^\circ$, Decl. = $22.01^\circ$, ICRS reference frame) for extracting them.

The Crab pulsar is a bright source; however, in the hard X-ray band ($>10$~keV) covered by NuSTAR, the count rate drops precipitously, and the spectrum is eventually dominated by instrumental background at the highest energies. Due to the short exposure times of many individual NuSTAR snapshots (often under 3~ks), fitting individual observations yields high parameter uncertainties. To ensure robust parameter estimation and sufficient signal-to-noise ratios, we implemented a grouping strategy. Sequential observations, sorted by Modified Julian Date (MJD), were combined into joint-fit batches until the cumulative exposure time reached a minimum threshold of 5.0~ks. After discarding two exposures shorter than 100~s, this strategy successfully yielded 43 independent measurement epochs with an average exposure of $\sim 7.9$~ks per batch.

All spectral modeling was performed utilizing \texttt{XSPEC} version 12.15 \citep{1996ASPC..101...17A}. Prior to spectral modeling, the spectra were grouped to a minimum of 30 counts per bin to ensure the validity of $\chi^2$ minimization statistics during fitting. For each $\ge 5$~ks batch, the on-pulse spectra was used as source, with off-pulse phase as background for both focal plane modules (FPMA and FPMB). The spectral channels were restricted to the strictly calibrated NuSTAR operating band of $3.0$--$79.0$~keV. We adopted the abundances of \citet{2000ApJ...542..914W} and the photoelectric cross-sections of \citet{1996ApJ...465..487V} for all interstellar absorption components.

Thus, we modeled the pulsed emission using a simple absorbed power law: \texttt{tbabs * powerlaw}. The interstellar hydrogen column density was fixed at the canonical value for the Crab, $N_{\rm H} = 0.36 \times 10^{22} \text{ cm}^{-2}$ \citep{2005SPIE.5898...22K}. The photon index ($\Gamma_{\rm X}$) and the power-law normalization were left as free parameters, though tied across all spectra within a given batch to compute a single, unified measurement for that epoch. A systematic uncertainty of 3\% is introduced in the spectral fits. This simple model provided very good fits across the dataset.

\subsection{The braking index}

The braking index $n$ is defined by the rotational kinematics of the pulsar as $n = \Omega \ddot{\Omega} / \dot{\Omega}^2$. We tracked the pulsar's rotational parameters utilizing the monthly timing ephemeris maintained by the Jodrell Bank Observatory\footnote{\url{https://www.jb.man.ac.uk/pulsar/crab.html}} \citep{1993MNRAS.265.1003L}. 
We estimate the second time derivative of the rotation frequency ($\ddot{\Omega}$) at each specific epoch by evaluating the difference between adjacent $\dot{\Omega}$ to yield rate of change. Statistical uncertainties were determined with error propagation.

\end{document}